\documentclass[pra,aps,groupaddress,prd]{revtex4}

\usepackage{amsfonts}
\usepackage{epsfig,amsmath,graphicx,amssymb,overpic,bm}
\usepackage{color,tikz,xcolor,extarrows}
\usetikzlibrary{arrows}

\usepackage{epsfig,amsmath,graphicx,amssymb,overpic}
\usepackage{bm}

\def\be{\begin{equation}}
\def\ee{\end{equation}}
\def\bee{\begin{eqnarray}}
\def\ene{\end{eqnarray}}
\def\bes{\begin{subequations}}
\def\ees{\end{subequations}}

\usepackage{epsfig,amsmath,graphicx,amssymb,overpic}
\def\be{\begin{equation}}
\def\ee{\end{equation}}
\def\bee{\begin{eqnarray}}
\def\ene{\end{eqnarray}}
\def\bes{\begin{subequations}}
\def\ees{\end{subequations}}
\def\no{\nonumber}

\def\d{\displaystyle}

\def\v{\vspace{0.1in}}
\def\no{{\nonumber}}

\begin{document}
\title{\large New integrable multi-L\'evy-index and mixed fractional nonlinear soliton hierarchies}
\author{Zhenya Yan}
\email[{\it Corresponding author}:\,\,]{zyyan@mmrc.iss.ac.cn}
\vspace{0.1in}
\affiliation{\small \vspace{0.1in} Key Laboratory of Mathematics Mechanization, Academy of Mathematics and Systems Science, Chinese Academy of Sciences, Beijing 100190, China \\
 School of Mathematical Sciences, University of Chinese Academy of Sciences, Beijing 100049, China}

\vspace{0.15in}
\baselineskip=12pt
\vspace{0.15in}
\begin{abstract}
\noindent {\bf Abstract.} In this letter, we present a simple and new idea to generate two types of novel integrable multi-L\'evy-index and mix-L\'evy-index (mixed) fractional nonlinear soliton hierarchies, containing multi-index and mixed fractional higher-order nonlinear Schr\"odinger (NLS) hierarchy, fractional complex modified Korteweg-de Vries (cmKdV) hierarchy, and fractional mKdV hierarchy. Their explicit forms can be given using the completeness of squared eigenfunctions. Moreover, we present their anomalous dispersion relations via their linearizations, and fractional multi-soliton solutions via the inverse scattering transform with matrix Riemann-Hilbert problems. These obtained fractional multi-soliton solutions may be useful to understand the related super-dispersion transports of nonlinear waves in multi-index fractional nonlinear media.

%

\vspace{0.1in}

\end{abstract}

\maketitle

\baselineskip=12pt

{\it Introduction.}---There exist many types of nonlinear equations in the fields of mathematical physics, in which an important type of nonlinear wave equations (NLWEs, called integrable equations) can be solved by the inverse scattering scattering (IST)~\cite{Gardner1967} to describe some nonlinear physical phenomena, such as the KdV equation, NLS equation, mKdV equation, Hirota equation, derivative NLS equation, and etc~\cite{ist1,ist2,rev1,rev2}. The partial derivatives in these integrable NLWEs are usual integer-order ones, that is, these nonlinear waves display the normal transports. In fact, fractional calculus (FC), as an extension of integer-order one, also plays an important role in many physical models with anomalous diffusions such as nanofluids, viscoelastic material, optics, Bose-Einstein condensates, geotechnical engineering, quantum mechanics, and  polymer science (see, e.g., Refs.~\cite{fc-book2,fc-book,pr20,prl87,west97,fc-rev,long15,barkai2012}), in which these fractional nonlinear physical models are usually not IST-integrable such that they were approximately solved using numerical methods (see, e.g., Refs.~\cite{fc-num,fls00,fc-book-num,wangprl2020,boris21} and reference therein).
More recently, based on nonlinear extensions of Riesz fractional derivative ($|\!-\!\partial_x^2|^{\epsilon},\, \epsilon\in (0, 1))$~\cite{Riesz,li19},  Ablowitz {\it et al} presented some new types of
IST-integrable fractional nonlinear soliton equations, such as the fractional NLS (fNLS), fractional KdV (fKdV), and fractional mKdV (fmKdV) equations~\cite{ab-prl22,ab22}. Moreover, the dynamics of new multi-soliton solutions was studied for integrable fractional higher-order Hirota, cmKdV and mKdV hierarchies by the IST with Riemann-Hilbert (RH) method~\cite{yanfnls-2022,yanfmkdv-2022}. It should be pointed out that all these integrable fractional nonlinear soliton equations contain {\it only single L\'evy index} $\epsilon\in (0, 1)$, that is, all spatial-derivative terms of these integrable fractional nonlinear soliton equations have the same operator $|\!-\!\partial_x^2|^{\epsilon}$. In fact, there exist some fractional models mixing  both normal and anomalous transports~\cite{fc-rev,wangprl2020,boris21}, however they are all non-integrable.

A natural problem is whether there are {\it new integrable  multi-L\'evy-index} ($\epsilon_\ell\in (0, 1),\, \ell=1,2,...)$ fractional nonlinear soliton equations (i.e., these operators $|\!-\!\partial_x^2|^{\epsilon_{\ell}}$ with two or more L\'evy indexes appearing in one equation) or even {\it integrable mix-L\'evy-index} (mixed) fractional nonlinear soliton equations (i.e., one equation admits not only fractional derivatives but also integer-order derivatives). To the best of our knowledge, no
multi-L\'evy-index or mixed fractional nonlinear wave equation was shown to be integrable before. In this paper, we would like to extend the combination of Riesz fractional derivative and IST to consider these issues such that some new classes of integrable multi-L\'evy-index and mixed fractional nonlinear soliton hierarchies can be found. Moreover, we give their anomalous dispersion relations, and multi-soliton solutions are obtained by using the IST with RH method.

{\it Multi-L\'evy-index and mixed fractional higher-order NLS hierarchy and anomalous dispersion relations}.---The second-order matrix ZS spectral problem~\cite{nls1} is
\bee
\label{lax-x}
\Phi_x=X\Phi,\quad X(x ,t; k)=-ik\sigma_3+U(x, t),
\quad U(x, t)=\begin{pmatrix} 0&q(x, t) \\ r(x, t)&0 \end{pmatrix},\quad
\sigma_3=\begin{pmatrix} 1&0\\ 0&-1 \end{pmatrix},
\ene
where $\Phi=\Phi(x,t;k)$ is a 2$\times$2 complex-valued matrix eigenfunction, $k\in \mathbb{C}$, and $r(x,t),\, q(x, t)$ denote the potentials. Usually, Eq.~(\ref{lax-x}) and associated evolution equations can generate the integrable integer-order AKNS hierarchy~\cite{ist1,ist2}. We here introduce the multi L\'evy indexes ($\epsilon_\ell\in (0, 1),\, \ell=2,3,...)$ and mix indexes ($\Pi_{\ell=2}^n\epsilon_\ell=0$ with $\sum_{\ell=2}^n\epsilon_\ell^2\not=0)$ to present the new integrable multi-L\'evy-index or mixed fractional higher-order NLS (fHONLS) hierarchy:
\bee \label{fnlss}
{\bf q}_t+\sigma_3{\cal N}_{h,\epsilon_\ell}(\widehat{\bf L}){\bf q}=0,\quad
{\cal N}_{h,\epsilon_\ell}(\widehat{\bf L})=\sum_{\ell=2}^n\alpha_\ell\chi_\ell|4\widehat{\bf L}^2|^{\epsilon_\ell}(2i\widehat{\bf L})^\ell,
\quad {\bf q}=(r(x, t),\, q(x, t))^{\rm T},\quad \epsilon_\ell\in (0, 1),
 \ene
 where $\alpha_\ell\in\mathbb{R}$ and $\chi_\ell=i\, (\ell$ even) and $\chi_\ell=1\, (\ell$ odd), and the operator $\widehat{\bf L}$ is defined as
 \bee  \widehat{\bf L}=\frac{1}{2i}\begin{pmatrix} \partial_x-2r\partial_{x-}^{-1}q &  2r\partial_{x-}^{-1}r \v\\
         -2q\partial_{x-}^{-1}q &  -\partial_x+2q\partial_{x-}^{-1}r \end{pmatrix},\quad
         \partial_x=\frac{\partial}{\partial x},\quad \partial_{x-}^{-1}=\int_{-\infty}^xdy.
\ene
Let $r=-\sigma q^*\, (\sigma=\pm 1)$, where the star denotes the complex conjugate, then Eq.~(\ref{fnlss}) can generate the new multi-L\'evy-index and mixed fHONLS hierarchy
\bee\label{hnls}
 i\begin{pmatrix} -\sigma q_t^* \\ q_t \end{pmatrix}+\sigma_3
 \left(\sum_{\ell=2}^n\alpha_\ell\chi_\ell i^{\ell+1}|4\widehat{\bf L}^2|^{\epsilon_\ell}(2\widehat{\bf L})^\ell\right)
 \begin{pmatrix}-\sigma q^* \\ q \end{pmatrix}=0.
\ene
As $\epsilon_\ell=0\, (\ell=2,3,...,n)$, Eq.~(\ref{hnls}) recovers the known integrable NLS hierarchy; as all $\epsilon_\ell\, (\ell=2,3,...,n)$ are equal, Eq.~(\ref{hnls}) recovers the integrable single-index fHONLS hierarchy, whereas
$\Pi_{\ell=2}^n\epsilon_\ell=0$ with $\sum_{\ell=2}^n\epsilon_\ell^2\not=0$, Eq.~\eqref{hnls} becomes the integrable
{\it new mixed} fractional fHONLS hierarchy containing not only the fractional derivatives (some nonzero indexes $\epsilon_{\ell_1}$) but also integer-order derivatives (other zero indexes $\epsilon_{\ell_2}$).

For example, we present the {\it new integrable multi-L\'evy-index and mixed} fHONLS equations from Eq.~\eqref{hnls}:
\begin{itemize}
\item [i)] Integrable single-index ($\epsilon_2$) fNLS equation ($n=2$)
\bee\label{fnls}
 i\begin{pmatrix} -\sigma q^*_t \\ q_t \end{pmatrix}-\alpha_2\sigma_3|4\widehat{\bf L}^2|^{\epsilon_2}
 \begin{pmatrix} -(\sigma q^*_{xx}+2|q|^2q^*) \vspace{0.05in} \\
q_{xx}+ 2\sigma|q|^2q \end{pmatrix}=0.
\ene
Notice that as $\epsilon_2=0$, Eq.~(\ref{fnls}) recovers the usual NLS equation.

\item [ii)] Integrable single-index ($\epsilon_3$) fcmKdV equation ($n=3$, $\alpha_2=0$)
\bee\label{m2}
 \begin{pmatrix} -\sigma q_t^* \\ q_t \end{pmatrix}-\alpha_3\sigma_3|4\widehat{\bf L}^2|^{\epsilon_3}\begin{pmatrix}
-(\sigma q^*_{xxx}+6|q|^2q^*_x) \vspace{0.05in} \\
q_{xxx}+6\sigma|q|^2q_x
\end{pmatrix}=0.
\ene
Notice that as $\epsilon_3=0$, Eq.~(\ref{m2}) recovers the usual cmKdV equation.

\item [iii)] New integrable two-L\'evy-index ($\epsilon_2,\, \epsilon_3$) fHirota equation ($n=3$)
\bee\label{fh}
 i\begin{pmatrix} -\sigma q_t^* \\ q_t \end{pmatrix}
 -\sigma_3\left[\alpha_2|4\widehat{\bf L}^2|^{\epsilon_2}\begin{pmatrix}
-(\sigma q^*_{xx}+2|q|^2q^*) \vspace{0.05in}\\ q_{xx}+ 2\sigma|q|^2q \end{pmatrix}
+i\alpha_3|4\widehat{\bf L}^2|^{\epsilon_3}\begin{pmatrix}
\sigma q^*_{xxx} + 6|q|^2q^*_x \vspace{0.05in}\\ q_{xxx}+6\sigma|q|^2q_x
\end{pmatrix} \right]=0.
\ene
Notice that as $\epsilon_2=\epsilon_3=0$, Eq.~\eqref{fh} recovers the Hirota equation~\cite{hirota}; as $\epsilon_2=\epsilon_3\in (0, 1)$, Eq.~\eqref{fh} recovers the integrable single-index fHirota equation~\cite{yanfnls-2022}; whereas $\epsilon_2=0$ or $\epsilon_3=0$, Eq.~\eqref{fh} becomes the two {\it new integrable mixed} fHirota equations:
\bee\label{fh1}
 i\begin{pmatrix} -\sigma q_t^* \\ q_t \end{pmatrix}
 -\sigma_3\left[\alpha_2|4\widehat{\bf L}^2|^{\epsilon_2}\begin{pmatrix}
-(\sigma q^*_{xx}+2|q|^2q^*) \vspace{0.05in}\\ q_{xx}+ 2\sigma|q|^2q \end{pmatrix}
+i\alpha_3\begin{pmatrix}
\sigma q^*_{xxx} + 6|q|^2q^*_x \vspace{0.05in}\\
q_{xxx}+6\sigma|q|^2q_x
\end{pmatrix} \right]
=0,
\ene
\vspace{-0.2in}\bee\label{fh2}
 i\begin{pmatrix} -\sigma q_t^* \\ q_t \end{pmatrix}
 -\sigma_3\left[\alpha_2\begin{pmatrix}
-(\sigma q^*_{xx}+2|q|^2q^*) \vspace{0.05in}\\ q_{xx}+ 2\sigma|q|^2q \end{pmatrix}
+i\alpha_3|4\widehat{\bf L}^2|^{\epsilon_3}\begin{pmatrix}
\sigma q^*_{xxx} + 6|q|^2q^*_x \vspace{0.05in}\\
q_{xxx}+6\sigma|q|^2q_x
\end{pmatrix} \right]
=0.
\ene

\item [iv)] New integrable two-L\'evy-index ($\epsilon_3,\, \epsilon_5$) fractional 3rd-5th-order cmKdV (f35cmKdV) equation ($n=5$, $\alpha_2=\alpha_4=0$)
\bee\label{fh35}
\begin{array}{l}
\begin{pmatrix} -\sigma q_t^* \vspace{0.05in} \\ q_t \end{pmatrix}
-\sigma_3\left[\alpha_3|4\widehat{\bf L}^2|^{\epsilon_3}\!\!\begin{pmatrix}
\sigma q^*_{xxx} + 6|q|^2q^*_x \vspace{0.05in}\\ q_{xxx}+6\sigma|q|^2q_x \end{pmatrix}  +\alpha_5|4\widehat{\bf L}^2|^{\epsilon_5}\!\!\begin{pmatrix}
 \sigma {\cal S}_5^* \vspace{0.05in}\\ {\cal S}_5 \end{pmatrix}\right]
=0
\end{array}
\ene
with $S_5[q]=q_{xxxxx}+ 10\sigma |q|^{2}q_{xxx}+10\sigma(q|q _{x}|^{2})_{x}+ 10 \sigma q^*(q_{x}^2)_{x}+30|q|^{4}q_{x}.$
Notice that as $\epsilon_3=\epsilon_5=0$, Eq.~\eqref{fh35} recovers the 35cmKdV equation; as $\epsilon_3=\epsilon_5\in (0, 1)$, Eq.~\eqref{fh35} recovers the integrable single-index f35cmKdV equation; whereas $\epsilon_3=0$ or $\epsilon_5=0$, Eq.~\eqref{fh35} becomes two {\it new integrable mixed} f35cmKdV equations:
\bee\label{fh35-1}
\begin{array}{l}
\begin{pmatrix} -\sigma q_t^* \vspace{0.05in} \\ q_t \end{pmatrix}
-\sigma_3\left[\alpha_3|4\widehat{\bf L}^2|^{\epsilon_3}\!\!\begin{pmatrix}
\sigma q^*_{xxx} + 6|q|^2q^*_x \vspace{0.05in}\\ q_{xxx}+6\sigma|q|^2q_x \end{pmatrix}  +\alpha_5\begin{pmatrix}
 \sigma {\cal S}_5^* \vspace{0.05in}\\ {\cal S}_5 \end{pmatrix}\right]
=0,
\end{array}
\ene
\vspace{-0.2in}\bee\label{fh35-2}
\begin{array}{l}
\begin{pmatrix} -\sigma q_t^* \vspace{0.05in} \\ q_t \end{pmatrix}
-\sigma_3\left[\alpha_3\begin{pmatrix}
\sigma q^*_{xxx} + 6|q|^2q^*_x \vspace{0.05in}\\ q_{xxx}+6\sigma|q|^2q_x \end{pmatrix}  +\alpha_5|4\widehat{\bf L}^2|^{\epsilon_5}\!\!\begin{pmatrix}
 \sigma {\cal S}_5^* \vspace{0.05in}\\ {\cal S}_5 \end{pmatrix}\right]
=0.
\end{array}
\ene

\item [v)] New integrable three-L\'evy-index ($\epsilon_2,\, \epsilon_3,\, \epsilon_4$) fractional fourth-order NLS (f4NLS) equation ($n=4$)
\bee\label{fh234}
\begin{array}{l}
i\begin{pmatrix}\! -\sigma q_t^* \vspace{0.05in} \\ q_t \!\!\end{pmatrix}
\!-\!\sigma_3\!\!\left[\alpha_2|4\widehat{\bf L}^2|^{\epsilon_2}\!\!\begin{pmatrix}\!
\!-\!(\sigma q^*_{xx}+2|q|^2q^*) \vspace{0.05in}\\ q_{xx}+ 2\sigma|q|^2q\!\!\end{pmatrix}
\!+\!i\alpha_3|4\widehat{\bf L}^2|^{\epsilon_3}\!\!\begin{pmatrix}
\sigma q^*_{xxx} + 6|q|^2q^*_x \vspace{0.05in}\\ q_{xxx}+6\sigma|q|^2q_x \end{pmatrix}
\!+\!\alpha_4|4\widehat{\bf L}^2|^{\epsilon_4}\!\!\begin{pmatrix}\!\!
 -\sigma {\cal S}_4^* \vspace{0.05in}\\ {\cal S}_4\!\! \end{pmatrix} \right]\!=\! 0
\end{array}
\ene
with ${\cal S}_4[q]=q_{xxxx}+8\sigma |q|^{2}q_{xx}+6|q|^{4}q +2\sigma q^{2}q_{xx}^*+4\sigma |q_{x}|^{2}q + 6 \sigma q^*q_{x}^{2}$.
Notice that as $\epsilon_j=0\, (j=2,3,4)$, Eq.~\eqref{fh234} recovers the integrable fourth-order NLS (4NLS) equation~\cite{lpd}; as $\epsilon_2=\epsilon_3=\epsilon_4\in (0, 1)$, Eq.~\eqref{fh234} recovers the integrable single-index f4NLS equation; whereas $\Pi_{j=2}^4\epsilon_j=0$ with $\sum_{j=2}^4\epsilon_j^2\not=0$, Eq.~\eqref{fh234} becomes six {\it new integrable mixed} f4NLS equations given by Eq.~\eqref{fh234} with $(\epsilon_2\epsilon_3\not=0,\, \epsilon_4=0),\, (\epsilon_2\epsilon_4\not=0,\, \epsilon_3=0),\,
(\epsilon_3\epsilon_4\not=0,\, \epsilon_2=0),\, (\epsilon_2\not=0,\, \epsilon_3=\epsilon_4=0),\, (\epsilon_3\not=0,\, \epsilon_2=\epsilon_4=0)$, or $(\epsilon_4\not=0,\, \epsilon_2=\epsilon_3=0)$.

\item [vi)] New integrable four-L\'evy-index ($\epsilon_2,\, \epsilon_3,\, \epsilon_4,\, \epsilon_5$) fractional fifth-order NLS (f5NLS) equation ($n=5$)
\bee\label{fh5}
\begin{array}{l}
i\begin{pmatrix} -\sigma q_t^* \vspace{0.05in} \\ q_t \end{pmatrix}
-\sigma_3\left[\alpha_2|4\widehat{\bf L}^2|^{\epsilon_2}\!\!\begin{pmatrix}
-(\sigma q^*_{xx}+2|q|^2q^*) \vspace{0.05in}\\ q_{xx}+ 2\sigma|q|^2q \end{pmatrix}
+i\alpha_3|4\widehat{\bf L}^2|^{\epsilon_3}\!\!\begin{pmatrix}
\sigma q^*_{xxx} + 6|q|^2q^*_x \vspace{0.05in}\\ q_{xxx}+6\sigma|q|^2q_x \end{pmatrix} \right.\vspace{0.05in}\\
\qquad\qquad\qquad\quad\,\, \left.+\alpha_4|4\widehat{\bf L}^2|^{\epsilon_4}\!\!\begin{pmatrix}
 -\sigma {\cal S}_4^* \vspace{0.05in}\\ {\cal S}_4 \end{pmatrix}
 +i\alpha_5|4\widehat{\bf L}^2|^{\epsilon_5}\!\!\begin{pmatrix}
 \sigma {\cal S}_5^* \vspace{0.05in}\\ {\cal S}_5 \end{pmatrix}\right]
=0.
\end{array}
\ene
Notice that as $\epsilon_j=0\, (j=2,3,4,5)$, Eq.~\eqref{fh5} recovers the integrable 5NLS equation; as $\epsilon_2=\epsilon_3=\epsilon_4=\epsilon_5\in (0, 1)$, Eq.~\eqref{fh5} recovers the integrable single-index f5NLS equation; whereas $\Pi_{j=2}^5\epsilon_j=0$ with $\sum_{j=2}^5\epsilon_j^2\not=0$, Eq.~\eqref{fh5} becomes fourteen new integrable mixed f5NLS equations.

\item [vii)] There are other new integrable multi-index and mixed fractional higher-order NLS equations for $n>5$.
\end{itemize}

We now use $q(x,t)\sim e^{i[kx-w_{h,\epsilon_\ell}(k)t]}$ to study the linearization of Eq.~(\ref{fnlss}) to yield the dispersion relation of the linear fHONLS hierarchy: $w_{h,\epsilon_\ell}(k)=i{\cal N}_{h,\epsilon_\ell}(-k/2)$, and further we can consider the linearization of the
multi-L\'evy-index fHONLS hierarchy (\ref{hnls})
\bee
 iq_t+\sum_{\ell=2}^n\alpha_\ell i^{\delta_\ell}|\!\!-\!\partial_x^2|^{\epsilon_\ell}q_{\ell x}=0,\quad
\delta_\ell=0,\,\, \ell=2m;\,\, \delta_\ell=1,\,\, \ell=2m+1,\,\, m\in \mathbb{N},
\ene
where $q_{\ell x}=\partial^\ell q/\partial x^\ell$,  $|\!\!-\!\partial_x^2|^{\epsilon}$ denotes the Riesz fractional derivative, such that  the anomalous dispersion relation and ${\cal N}_{h,\epsilon_\ell}(k)$ are given as
\bee\label{Ng}
 w_{h,\epsilon_\ell}(k)=-\sum_{\ell=2}^n\alpha_\ell i^{\delta_\ell+\ell}k^\ell|k^2|^{\epsilon_\ell},\quad
 {\cal N}_{h,\epsilon_\ell}(k)=-iw_{h,\epsilon_\ell}(-2k)=\sum_{\ell=2}^n\alpha_\ell i^{\delta_\ell+\ell+1}(-2k)^\ell|4k^2|^{\epsilon_\ell}.
 \ene

{\it Multi-L\'evy-index and mixed fractional higher-order mKdV hierarchy and anomalous dispersion relations}.---Let $r=-\sigma q\, \, (\sigma=\pm 1)$ with $q(x,t)\in\mathbb{R}[x,t]$ in the spectral problem \eqref{lax-x}, then, similarly, we
can give the new integrable multi-L\'evy-index and mixed fractional higher-order mKdV (fHOmKdV) hierarchy
\bee \label{fmkdvh}
q_t+{\cal M}_{h,\epsilon_\ell}(\widehat{\cal L})q_x=0,\quad {\cal M}_{h,\epsilon_\ell}(\widehat{\cal L})=\sum_{\ell=1}^n\alpha_{2\ell+1}(-\widehat{\cal L})^\ell|\widehat{\cal L}|^{\epsilon_{2\ell+1}},\quad \epsilon_\ell\in (0, 1),
 \ene
where $\alpha_{2\ell+1}\in\mathbb{R},\, \widehat{\cal L}=-\partial_x^2-4\sigma q^2-4\sigma q_x\partial_{x-}^{-1}q$, which can also be rewritten as
\bee\label{fmkdvh2}
 q_t+\sum_{\ell=1}^n\alpha_{2\ell+1}|\widehat{\cal L}|^{\epsilon_{2\ell+1}}\partial_x{\cal F}_{2\ell+1}[q(x,t)]=0,
\ene
where ${\cal F}_{2\ell+1}[q(x,t)]$'s are given as~\cite{hmkdv1}
 \begin{align}
&{\cal F}_3[q(x,t)]=q_{xx}+2\sigma q^3, \no \v\\ 
&{\cal F}_5[q(x,t)]=q_{xxxx}+10\sigma (q^2q_{xx}+qq_x^2)+6q^5, \no \v\\
&{\cal F}_7[q(x,t)]=q_{6x}+14\sigma (q^2q_{xxxx}+4qq_xq_{xxx}+3qq_{xx}^2+5q_x^2q_{xx}+5q^4q_{xx}+10q^3q_x^2)+20q^7,... \no
\end{align}
As $\epsilon_{2\ell+1}=0\, (\ell=1,2,...,n)$, Eq.~\eqref{fmkdvh} or (\ref{fmkdvh2}) recovers the known integrable integer-order mKdV hierarchy; as all $\epsilon_\ell\, (\ell=2,3,...,n)$ are equal, Eq.~\eqref{fmkdvh} or (\ref{fmkdvh2}) recovers the integrable  single-index fHOmKdV hierarchy~\cite{yanfmkdv-2022}, whereas $\Pi_{\ell=1}^n\epsilon_{2\ell+1}=0$ with $\sum_{\ell=1}^n\epsilon_{2\ell+1}^2\not=0$, Eq.~\eqref{fmkdvh} or (\ref{fmkdvh2})  becomes the
{\it new integrable mixed} fractional higher-order mKdV hierarchy.

In particular, we have some examples on the new multi-index and mixed fHOmKdV hierarchy:
\begin{itemize}
\item [i)] Integrable single-index ($\epsilon_3$) fractional mKdV (fmKdV) equation ($n=1$)
 \bee\label{fmkdv}
 q_t+\alpha_3|\widehat{\cal L}|^{\epsilon_3}(q_{xxx}+6\sigma q^2q_x)=0.
\ene
Notice that as $\epsilon_3=0$, Eq.~(\ref{fmkdv}) recovers the usual mKdV equation.

\item [ii)] Integrable single-index ($\epsilon_5$) fractional fifth-order mKdV (f5mKdV) equation ($n=2,\,\alpha_3=0$)
 \bee\label{fmkdv5}
  u_t+\alpha_5|\widehat{\cal L}|^{\epsilon_5}\!\left(q_{xxxxx}+10\sigma(q^2q_{xx}+qq_x^2)_x+30q^4q_x\right)=0.
\ene
Notice that as $\epsilon_5=0$, Eq.~(\ref{fmkdv5}) recovers the usual 5mKdV equation.

\item [iii)] New integrable two-L\'evy-index ($\epsilon_3,\, \epsilon_5$) fractional 3rd-5th-order mKdV (f35mKdV) equation ($n=2$)
 \bee\label{fmkdv5}
   q_t+\alpha_3|\widehat{\cal L}|^{\epsilon_3}(q_{xxx}+6\sigma q^2q_x)
   +\alpha_5|\widehat{\cal L}|^{\epsilon_5}\!\left(q_{xxxxx}+10\sigma(q^2q_{xx}+qq_x^2)_x+30q^4q_x\right)=0.
\ene
Notice that as $\epsilon_3=\epsilon_5=0$, Eq.~\eqref{fmkdv5} recovers the 35mKdV equation; as $\epsilon_3=\epsilon_5\in (0, 1)$,
Eq.~\eqref{fmkdv5} recovers the single-index f35mKdV equation; whereas $\epsilon_3=0$ or $\epsilon_5=0$,
Eq.~\eqref{fmkdv5} becomes two {\it new integrable mixed} f35mKdV equations:
\bee\label{fmkdv5-1}
 \begin{array}{l} q_t+\alpha_3|\widehat{\cal L}|^{\epsilon_3}(q_{xxx}+6\sigma q^2q_x)+\alpha_5\!\left(q_{xxxxx}+10\sigma(q^2q_{xx}+qq_x^2)_x+30q^4q_x\right)=0, \v\\
  q_t+\alpha_3(q_{xxx}+6\sigma q^2q_x)+\alpha_5|\widehat{\cal L}|^{\epsilon_5}\!\left(q_{xxxxx}+10\sigma(q^2q_{xx}+qq_x^2)_x+30q^4q_x\right)=0.
\end{array}
\ene

\item [iv)] New integrable three-L\'evy-index ($\epsilon_3,\, \epsilon_5,\, \epsilon_7$) fractional 3rd-5th-7th-order mKdV (f357mKdV) equation ($n=3)$
 \bee\label{fmkdv7}
   q_t+\alpha_3|\widehat{\cal L}|^{\epsilon_3}(q_{xxx}+6\sigma q^2q_x)+\alpha_5|\widehat{\cal L}|^{\epsilon_5}(q_{xxxxx}+10\sigma(q^2q_{xx}+qq_x^2)_x+30q^4q_x)
   +\alpha_7|\widehat{\cal L}|^{\epsilon_7}\partial_x{\cal F}_7[q]=0.
\ene
Notice that as $\epsilon_{2\ell+1}=0\, (\ell=1,2,3)$, Eq.~\eqref{fmkdv7} recovers the 7th-order mKdV equation; as $\epsilon_3=\epsilon_5=\epsilon_7\in (0, 1)$, Eq.~\eqref{fmkdv7} recovers the integrable single-index f357mKdV equation; whereas
$\Pi_{\ell=1}^3\epsilon_{2\ell+1}=0$ with $\sum_{\ell=1}^3\epsilon_{2\ell+1}^2\not=0$, Eq.~\eqref{fmkdv7} becomes six {\it new integrable mixed} fractional 7th-order mKdV equations  given by Eq.~\eqref{fmkdv7} with $(\epsilon_3\epsilon_5\not=0,\, \epsilon_7=0),\, (\epsilon_3\epsilon_7\not=0,\, \epsilon_5=0),\, (\epsilon_5\epsilon_7\not=0,\, \epsilon_3=0),\, (\epsilon_3\not=0,\, \epsilon_5=\epsilon_7=0),\, (\epsilon_5\not=0,\, \epsilon_3=\epsilon_7=0)$, or $(\epsilon_7\not=0,\, \epsilon_3=\epsilon_5=0)$.

\item [v)] There are other new integrable multi-L\'evy-index and mixed fractional higher-order mKdV equations for $n>3$.
\end{itemize}

Here we apply $q(x,t)\sim e^{i[kx-\omega_{h,\epsilon_\ell}(k)t]}$ to consider the linearization of Eq.~(\ref{fmkdvh}) to yield the dispersion relation of the multi-index linear fHOmKdV hierarchy: $\omega_{h,\epsilon_\ell}(k)=k{\cal M}_{h,\epsilon_\ell}(k^2)$, and further we can consider the linearization of Eq.~\eqref{fmkdvh2} such that the anomalous dispersion relation and ${\cal M}_{h,\epsilon_\ell}(k^2)$ are
\bee \label{Ngk}
 w_{h,\epsilon_\ell}(k)=\sum_{\ell=1}^n\alpha_{2\ell+1}(-1)^\ell k^{2\ell+1}|k^2|^{\epsilon_\ell},\quad
  {\cal M}_{h,\epsilon_\ell}(k^2)=\frac{w_{h,\epsilon_\ell}(k)}{k}=\sum_{\ell=1}^n\alpha_{2\ell+1}(-k^2)^{\ell}|k^2|^{\epsilon_\ell}.
 \ene

{\it Completeness of squared eigenfunctions and explicit multi-L\'evy-index and mixed fHONLS and fHOmKdV hierarchies}.---For the given ZS-AKNS spectral problem (\ref{lax-x}), we now consider the time evolution of the matrix eigenfunction $\Phi(x,t; k)$ as
 \bee\label{lax-t}
\begin{array}{ll}
\Phi_t=T\Phi, & \quad
T(x,t; k)=\begin{pmatrix}
  T_a(x,t; k) & T_b(x,t; k) \vspace{0.05in}\\  T_c(x,t; k) & -T_a(x,t; k)
  \end{pmatrix},
\end{array}
\ene
where, in general, these functions $T_{a,b,c}(x,t; k)$ can not be explicitly given for the multi-index and mixed fractional higher-order NLS and mKdV hierarchies. However we here assume these conditions $T_{b,c}(x,t; k)\to 0$ and
$T_a(x,t; k)\to \frac12 {\cal N}_{h,\epsilon_\ell}(k)$ and $ik{\cal M}_{h,\epsilon_\ell}(4k^2)$ with ${\cal N}_{h,\epsilon_\ell}(k),\, {\cal M}_{h,\epsilon_\ell}(k)$ given by Eqs.~(\ref{Ng}) and
(\ref{Ngk}) for the multi-index and mixed fHONLS and fHOmKdV hierarchies, respectively, as $x\to \pm \infty$, (i.e., $q(x,t)\to 0)$. Therefore, for the zero-boundary condition $q(x,t)\in L^1(\mathbb{R}^{\pm})$ with $r=-q^*$ and $r=-q$ for the focusing multi-index and mixed fHONLS and fHOmKdV hierarchies, respectively, we consider the corresponding asymptotic problems $(x\to \pm \infty)$ of the spectral problem (\ref{lax-x}) and time part (\ref{lax-t}) to yield
\bee \label{bianjie0}
{\rm fHONLS:}\quad \Phi_{\pm}^{(\mathbb{C})}(x,t;k)=\phi_{\pm}^{(\mathbb{C})}(x,t;k)e^{(-ikx+\frac12{\cal N}_{h,\epsilon_\ell}(k)t)\sigma_3}\to \mathbb{I},\quad x\to\pm\infty, \\
{\rm fHOmKdV:}\quad \Phi_{\pm}^{(\mathbb{R})}(x,t;k)=\phi_{\pm}^{(\mathbb{R})}(x,t;k)e^{-ik[x-{\cal M}_{h,\epsilon_\ell}(4k^2)t]\sigma_3}\to \mathbb{I},\quad x\to\pm\infty
\ene
with $\phi_{\pm}^{(\mathbb{C}/\mathbb{R})}(x,t;k)=\mathbb{I}+\int_{\pm\infty}^xdye^{-ik(x-y)\hat{\sigma}_3}U^{(\mathbb{C}/\mathbb{R})}(y,t)\phi_{\pm}(y,t;k),\, e^{\hat{\sigma}_3}U^{(\mathbb{C}/\mathbb{R})}=e^{\sigma_3}U^{(\mathbb{C}/\mathbb{R})}e^{-\sigma_3}$ with $U^{(\mathbb{C}/\mathbb{R})}=U|_{r=-q^*/r=-q}.$

Since $\Phi_{\pm}(x, t; k)$ are both fundamental solutions of the spectral problem, thus they have the linear relation
\bee\label{sr}
 \Phi_+^{(\mathbb{C}/\mathbb{R})}(x,t;k)=\Phi_-(x,t;k)^{(\mathbb{C}/\mathbb{R})}S^{(\mathbb{C}/\mathbb{R})}(k),\quad S^{(\mathbb{C}/\mathbb{R})}(k)=\left(s_{ij}^{(\mathbb{C}/\mathbb{R})}(k)\right)_{2\times 2}, \quad  k\in\mathbb{R}.
 \ene
As a result, one has $s_{ij}^{(\mathbb{C}/\mathbb{R})}(k)=(-1)^{i+1}\big|\Phi_{+j}^{(\mathbb{C}/\mathbb{R})}(x, t; k),\,\, \Phi_{-(3-i)}^{(\mathbb{C}/\mathbb{R})}(x, t; k)\big|,\,\, i,j=1,2$, where  $\Phi_{\pm}^{(\mathbb{C}/\mathbb{R})}(x, t; k)=(\Phi_{\pm 1}^{(\mathbb{C}/\mathbb{R})},\, \Phi_{\pm 2}^{(\mathbb{C}/\mathbb{R})})$. Since the eigenfunctions $\Phi_{-1, +2}^{(\mathbb{C}/\mathbb{R})}$ ($\Phi_{-2, +1}^{(\mathbb{C}/\mathbb{R})}$) can be extended analytically to $\mathbb{C}^{+}$ ($\mathbb{C}^{-}$), and continuously to $\mathbb{C}^{+}\cup \mathbb{R}$ ($\mathbb{C}^{-}\cup \mathbb{R}$), thus $s_{11}^{(\mathbb{C}/\mathbb{R})}(k)$ ($s_{22}^{(\mathbb{C}/\mathbb{R})}(k)$) in $k\in\mathbb{R}$ can be extended analytically to $\mathbb{C}^{-}$ ($\mathbb{C}^{+}$), and continuously to $\mathbb{C}^{-}\cup \mathbb{R}$ ($\mathbb{C}^{+}\cup \mathbb{R}$), whereas another two scattering coefficients $s_{12}^{(\mathbb{C}/\mathbb{R})}(k)$ and $s_{21}^{(\mathbb{C}/\mathbb{R})}(k)$ can not be analytically continued away from $\mathbb{R}$. Moreover, the potential function is required not to possess spectral singularities in order to solve the matrix RH problem of the inverse scattering~\cite{Zhou1989}.

{\it Case 1.}\, For the multi-index and mixed fHONLS hierarchy, an adjoint of $\widehat{\bf L}$ is of the form
\bee {\bf L}=\frac{1}{2i}\begin{pmatrix}
      -\partial_x-2q\partial_{x+}^{-1}r &  -2q\partial_{x+}^{-1}q \v\\
         2r\partial_{x+}^{-1}r &  \partial_x+2r\partial_{x+}^{-1}q \end{pmatrix}, \qquad \partial_{x+}^{-1}=\int_x^{\infty} dy.
\ene
It follows from the ZS spectral problem \eqref{lax-x} with $r=-q^*$ that the eigenfunctions ${\Psi}_{+j},\, \widehat{\Psi}_{-j}$ of ${\bf L}$ and $\widehat{\bf L}$ with eigenvalue $k$
are complete~\cite{akns,kaup76}, where ${\bf L}{\Psi}_{+j}=k{\Psi}_{+j},\, \widehat{\bf L}\widehat{\Psi}_{-j}=k\widehat{\Psi}_{-j}, \, (j=1,2)$ with
$\Psi_{+j}=(\Phi^{(\mathbb{C})2}_{+j1},\,  \Phi_{+j2}^{(\mathbb{C})2})^{\rm T}, \, \widehat{\Psi}_{-j}=(\Phi_{-j2}^{(\mathbb{C})2},\,  -\Phi_{-j1}^{(\mathbb{C})2})^{\rm T}$, where $\Phi_{\pm j}^{(\mathbb{C})}=(\Phi_{\pm j1}^{(\mathbb{C})}, \Phi_{\pm j2}^{(\mathbb{C})})^{\rm T}.$ Similarly, ${\cal N}_{h,\epsilon_\ell}(\widehat{\bf L})\widehat{\Psi}_{-j}={\cal N}_{h,\epsilon_\ell}(k)\widehat{\Psi}_{-j},\, j=1,2.$  Based on the completeness of the eigenfunctions $\widehat{\Psi}_{-j}$'s, one can act the operator  ${\cal N}_{h,\epsilon_\ell}(\widehat{\bf L})$ on some sufficiently smooth and decaying vector function ${\bf p}(x)=(p_1(x),\, p_2(x))^{\mathrm{T}}$ to yield
\bee
\label{jifen1}
{\cal N}_{h,\epsilon_\ell}(\widehat{\bf L}){\bf v}(x)=\frac{1}{\pi}\sum\limits_{j=1}^2\int_{\Gamma_{\infty}^{(j)}}dk
\hat s_{j}(k){\cal N}_{h,\epsilon_\ell}(k)\int_{\mathbb{R}}{\cal K}_{j}(x,y; k){\bf p}(y)dy,
\ene
where ${\cal K}_j(x,y; k)=\widehat{\Psi}_{-j}(x,k)\Psi_{+(3-j)}^{\mathrm T}(y,k),\, \hat s_{(3-j)}(k)=(-1)^{j+1}/s_{jj}^{(\mathbb{C})2}(k)$,
and $\Gamma_{\infty}^{(j)}=\lim\limits_{\mathbb{R}\to\infty}\Gamma_{\mathbb{R}}^{(j)}$ with
$\Gamma_{\mathbb{R}}^{(1)}$ ($\Gamma_{\mathbb{R}}^{(2)}$) being the semicircular contour in the upper (lower) half plane evaluated from $-\mathbb{R}$ to $\mathbb{R}$.

Then, it follows from ${\cal N}_{h,\epsilon_\ell}(\widehat{\bf L})$ defined by Eq.~\eqref{fnlss} acting on  ${\bf q}=(-q^*, q)^T$ (here we consider the focusing case $\sigma=1$) and Eq.~\eqref{jifen1} that the multi-L\'evy-index and mixed fHONLS hierarchy \eqref{hnls} can be explicitly written as
\bee \label{fhirota-exp}
 iq_t+\frac{1}{\pi}\sum\limits_{\ell=2}^n\sum\limits_{j=1}^2 \int_{\Gamma_{\infty}^{(j)}}dk|4k^2|^{\epsilon_\ell}
 \hat s_j(k) \int_{\mathbb{R}}{\cal K}_{j}^{(2)}(x,y;k)\Big(2i\widehat{\bf L}(y, t)\Big)^\ell \begin{pmatrix}-q^*(y,t) \vspace{0.05in} \\ q(y,t) \end{pmatrix} dy=0,
  \ene
where ${\cal K}_j^{(2)}(x,y;k)=-\Phi_{-j1}^{(\mathbb{C})2}(x,k)\Big(\Phi_{+(3-j)1}^{(\mathbb{C})2}(y,k),\, \Phi_{+(3-j)2}^{(\mathbb{C})2}(y,k)\Big)$.

For example, we have the new two-L\'evy-index ($\epsilon_2, \epsilon_3$) fHirota equation
\bee \label{fhirota-exp}
 iq_t\!+\!\frac{1}{\pi}\!\sum\limits_{j=1}^2 \int_{\Gamma_{\infty}^{(j)}}dk\hat s_j(k)\!\!\int_{\mathbb{R}}{\cal K}_{j}(x,y;k)
\!\!\left[\alpha_2|4k^2|^{\epsilon_2}\!\!\begin{pmatrix}\!-q^*_{xx}\!-\! 2|q|^2q^* \vspace{0.05in}\\ q_{xx} \!+\! 2|q|^2q \!\end{pmatrix}
\!+\!i\alpha_3|4k^2|^{\epsilon_3}\!\!\begin{pmatrix}\!  q^*_{xxx} \!+\! 6|q|^2q^*_x \vspace{0.05in}\\ q_{xxx}\!+\! 6|q|^2q_x \!\end{pmatrix}\!\right] dy=0,
  \ene
and two new mixed fHirota equations containing both fractional derivative and pure integer derivative terms:
\bee \label{fhirota-exp2}
 iq_t\!+\!\frac{1}{\pi}\!\sum\limits_{j=1}^2 \int_{\Gamma_{\infty}^{(j)}}dk\hat s_j(k)\!\!\int_{\mathbb{R}}{\cal K}_{j}(x,y;k)
\!\!\left[\alpha_2|4k^2|^{\epsilon_2}\!\!\begin{pmatrix}\!-q^*_{xx}\!-\! 2|q|^2q^* \vspace{0.05in}\\ q_{xx} \!+\! 2|q|^2q \!\end{pmatrix}
\!+\!i\alpha_3\!\!\begin{pmatrix}\!  q^*_{xxx} \!+\! 6|q|^2q^*_x \vspace{0.05in}\\ q_{xxx}\!+\! 6|q|^2q_x \!\end{pmatrix}\!\right] dy=0,
  \ene
\vspace{-0.2in}
\bee \label{fhirota-exp3}
 iq_t\!+\!\frac{1}{\pi}\!\sum\limits_{j=1}^2 \int_{\Gamma_{\infty}^{(j)}}dk\hat s_j(k)\!\!\int_{\mathbb{R}}{\cal K}_{j}(x,y;k)
\!\!\left[\alpha_2\begin{pmatrix}\!-q^*_{xx}\!-\! 2|q|^2q^* \vspace{0.05in}\\ q_{xx} \!+\! 2|q|^2q \!\end{pmatrix}
\!+\!i\alpha_3|4k^2|^{\epsilon_3}\!\!\begin{pmatrix}\!  q^*_{xxx} \!+\! 6|q|^2q^*_x \vspace{0.05in}\\ q_{xxx}\!+\! 6|q|^2q_x \!\end{pmatrix}\!\right] dy=0.
  \ene

{\it Case 2.}\, For the multi-L\'evy-index and mixed fHOmKdV hierarchy, it follows from the spectral problem \eqref{lax-x} with $r=-q$ that $\widehat{\cal L}\psi_{-1}=4k^2 \psi_{-1},\,\, {\cal L}\psi_{+2}=4k^2 \psi_{+2},$ where ${\cal L}=-\partial_x^2-4q^2-4q\partial_{x+}^{-1}q_y,\,\,\, \partial_{x+}^{-1}=\int_x^{\infty}dy$,\,\,\,
$\psi_{\pm j}=\Phi_{-j1}^{(\mathbb{R})2} \pm \Phi_{-j2}^{(\mathbb{R})2}\, (j=1,2).$
Based on the completeness of squared scalar eigenfunctions~\cite{kaup76,ab22}, one can act
${\cal M}_{h,\epsilon_\ell}(\widehat{\cal L})$ on some sufficiently smooth and decaying scalar function $g(x)$ to  yield
\bee\label{jifen12}
{\cal M}_{h,\epsilon_\ell}(\widehat{\cal L})g(x)=\frac{1}{\pi}\int_{\Gamma_{\infty}}dk
{\cal M}_{h,\epsilon_\ell}(k^2)s_{11}^{(\mathbb{R})-2}(k)\int_{\mathbb{R}}\psi_{-1}(x,k)\psi_{+2}(y,k)g(y)dy,
\ene
where $\Gamma_{\infty}=\lim_{{\mathbb R}\to \infty}\Gamma_{{\mathbb R}}$ with $\Gamma_{{\mathbb R}}$ being the semicircular contour in the upper half plane evaluated from $-{\mathbb R}$ to ${\mathbb R}$.

For the defined ${\cal M}_{h,\epsilon_\ell}(\widehat{\cal L})$ in Eq.~\eqref{fmkdvh}, one has the multi-index and mixed fHOmKdV hierarchy as
\bee\label{fmkdv-exp}
q_t+\frac{1}{\pi}\int_{\Gamma_{\infty}}\!\! dk\!\int_{\mathbb{R}}{\cal P}(x,y; k)\Big(\sum_{\ell=1}^n\alpha_{2\ell+1}|4k^2|^{\epsilon_{2\ell+1}}(-\widehat{\cal L}(y,t))^\ell q_y\Big) dy=0,
\ene
or
\bee\label{fmkdv-exp2g}
q_t+\frac{1}{\pi}\int_{\Gamma_{\infty}}\!\! dk\!\int_{\mathbb{R}}{\cal P}(x,y;k)
\Big(\sum_{\ell=1}^n\alpha_{2\ell+1}|4k^2|^{\epsilon_{2\ell+1}}\partial_y{\cal F}_{2\ell+1}[q(y,t)]\Big) dy=0,
\ene
where ${\cal P}(x,y; k)=\psi_{-1}(x,k)\psi_{+2}(y,k)/s_{11}^{(\mathbb{R})2}(k),\, \widehat{\cal L}(y,t)=-\partial_y^2-4q^2(y,t)-4q_y(y,t)\partial_{y-}^{-1}q(y, t)$ with $\partial_{y-}^{-1}=\int_{-\infty}^yds$.

For example, we have the new two-L\'evy-index ($\epsilon_3, \epsilon_5$) f35mKdV equation ($n=2$)
\bee\label{f35mkdv}
q_t+\frac{1}{\pi}\int_{\Gamma_{\infty}}\!\!dk\!\!\int_{\mathbb{R}}{\cal P}(x,y; k)
\Big[\alpha_3|4k^2|^{\epsilon_3}(q_{yyy}+6q^2q_y)
 \!+\!\alpha_5|4k^2|^{\epsilon_5}\!\left(q_{yyyyy}+10(q^2q_{yy}+qq_y^2)_y+30q^4q_y\right)\!\Big]dy=0.
\ene
and new three-L\'evy-index ($\epsilon_3, \epsilon_5, \epsilon_7$) f357mKdV equation ($n=3$)
\bee\label{f357mkdv}
q_t\!+\!\frac{1}{\pi}\int_{\Gamma_{\infty}}\!\!dk\!\!\int_{\mathbb{R}}{\cal P}(x,y;k)\!
\Big[\alpha_3|4k^2|^{\epsilon_3}(q_{yyy}+6q^2q_y)\!+\!\alpha_5|4k^2|^{\epsilon_5}\partial_y{\cal F}_5[q(y,t)
\!+\!\alpha_7|4k^2|^{\epsilon_7}\partial_y{\cal F}_7[q(y,t)]\Big]dy=0.
\ene
In particular, we have the new mixed f35mKdV equations ($n=2$)
\bee\label{f35mkdv2}
q_t+\frac{1}{\pi}\int_{\Gamma_{\infty}}\!\!dk\!\!\int_{\mathbb{R}}{\cal P}(x,y; k)
\Big[\alpha_3|4k^2|^{\epsilon_3}(q_{yyy}+6q^2q_y)
 \!+\!\alpha_5\!\left(q_{yyyyy}+10(q^2q_{yy}+qq_y^2)_y+30q^4q_y\right)\!\Big]dy=0,
\ene
\vspace{-0.2in}
\bee\label{f35mkdv3}
q_t+\frac{1}{\pi}\int_{\Gamma_{\infty}}\!\!dk\!\!\int_{\mathbb{R}}{\cal P}(x,y; k)
\Big[\alpha_3(q_{yyy}+6q^2q_y)
 \!+\!\alpha_5|4k^2|^{\epsilon_5}\!\left(q_{yyyyy}+10(q^2q_{yy}+qq_y^2)_y+30q^4q_y\right)\!\Big]dy=0,
\ene
and new six mixed f357mKdV equations ($n=3$):
\bee\label{f357mkdv2}
q_t+\frac{1}{\pi}\int_{\Gamma_{\infty}}dk\int_{\mathbb{R}}{\cal P}(x,y; k)
\Big[\alpha_3|4k^2|^{\epsilon_3}(q_{yyy}+6q^2q_y)+\alpha_5|4k^2|^{\epsilon_5}\partial_y{\cal F}_5[q(y,t)
+\alpha_7\partial_y{\cal F}_7[q(y,t)]\Big]dy=0,
\ene
and Eq.~\eqref{f357mkdv} with $(\epsilon_3\epsilon_7\not=0,\, \epsilon_5=0),\, (\epsilon_5\epsilon_7\not=0,\, \epsilon_3=0),\, (\epsilon_3\not=0,\, \epsilon_5=\epsilon_7=0),\, (\epsilon_5\not=0,\, \epsilon_3=\epsilon_7=0)$, or $(\epsilon_7\not=0,\, \epsilon_3=\epsilon_5=0)$.

\v {\it Fractional multi-soliton solutions via the IST.}---For the given initial condition $q(x, 0)$ with sufficient smoothness and decay, we can use the IST with the Riemann-Hilbert method to find the fractional multi-soliton solutions of the focusing ($\sigma=1$) multi-index and mixed fHONLS hierarchy \eqref{hnls}
\bee\label{recons}
q_{\tiny{\rm fHONLS}}(x, t)=-2i\dfrac{\left|\begin{matrix} \mathbb{I}_N+M & {\bf a}^{\rm T} \vspace{0.02in}\\ {\bf b} &0 \end{matrix}\right|}{|\mathbb{I}_N+M|},
\ene
where $\mathbb{I}_N$ is an $N$th-order unit matrix, ${\bf a}=(1,1,..,1),\, {\bf b}=(b_j)_{1\times N}=(b_1, b_2,..,b_N),\, b_n=B_+[k_n]\mathrm{e}^{2i\theta_{\epsilon}(x, t; k_n)},\, M=(M_{(n,n_2)})_{N\times N},\, M_{(n,n_2)}=\sum_{n_1=1}^{N}B_-[k_{n_1}^*]B_+[k_{n_2}]\,
\mathrm{e}^{-2i\theta_{\epsilon}(x, t; k_{n_1}^*)+2i\theta_{\epsilon}(x, t; k_{n_2})}/
[(k_n-k_{n_1}^*)(k_{n_2}-k_{n_1}^*)],\, n=1,2,...,N$ with $\theta_{\epsilon}(x,t; k)=-kx+\frac12\left(\sum_{\ell=2}^N\alpha_\ell i^{\delta_\ell+\ell}(-2k)^\ell|4k^2|^{\epsilon_\ell}\right)t$ and $\{\{k_n,\, k_n^*\} | s_{22}^{(\mathbb{C})}(k_n)=0,\, k_n\in \mathbb{C}^+\}$ is the discrete spectrum set and $B_+[k_n]$'s are constants with $B_-[k_n^*]=-B_+^*[k_n]$.
Notice that due to the limited space, we will present the detailed analysis of the fractional multi-soliton solutions via the IST with Riemann-Hilbert method in another work to be published elsewhere~\cite{yan2022}.

For the case of $N=1$, Let $k_1=\xi+i\eta$ and $B_+[k_1]=b$ with $b,\,\xi,\, \eta\in\mathbb{R}$ with $b\eta\not=0$, then we have the fractional one-soliton solution of the multi-index and mixed fHONLS hierarchy \eqref{hnls} as
\bee\label{solu1}
 q_{\rm fHONLS,1}(x,t)\!=\!2\eta\,{\rm sech}\!\left[2\eta x-\left(\sum_{\ell=2}^N\alpha_\ell i^{\delta_\ell+\ell}{\rm Im}[(-2k)^\ell]|4k_1^2|^{\epsilon_\ell}\right)t\!-\!\ln\left|\frac{2\eta}{b}\right|\right] \no \\
 \times e^{i[-2\xi x+\left(\sum_{\ell=2}^N\alpha_\ell i^{\delta_\ell+\ell}{\rm Re}[(-2k)^\ell]|4k_1^2|^{\epsilon_\ell}\right)t+\pi/2+{\rm Arg}(b/2\eta)]},
\ene
which admits the multi L\'evy indexes ($\epsilon_\ell\, (\ell=2,3,...,n$) and displays the anomalous dispersion. The wave and phase velocities are related to the multi L'evy indexes, given by 
\bee \no
\begin{array}{l}
v_w=\dfrac{1}{2\eta}\d\sum_{\ell=2}^N\alpha_\ell i^{\delta_\ell+\ell}{\rm Im}[(-2k)^\ell]|4k_1^2|^{\epsilon_\ell}, \quad v_p=\dfrac{1}{2\xi}\d\sum_{\ell=2}^N\alpha_\ell i^{\delta_\ell+\ell}{\rm Re}[(-2k)^\ell]|4k_1^2|^{\epsilon_\ell}.
\end{array}
\ene

Similarly, we can also use the IST with the Riemann-Hilbert method to find the fractional multi-soliton solutions of the focusing ($\sigma=1$) multi-index and mixed fHOmKdV hierarchy \eqref{fmkdvh2} as
\bee\label{mkdv}
q_{\rm fHOmKdV}(x, t)=-2i\dfrac{\left|\begin{matrix} 0 & \hat{\bf a} \vspace{0.02in}\\ \hat {\bf b}^{\rm T} & \Omega \end{matrix}\right|}{|\Omega|},
\ene
where $\hat{\bf a}=(a_{11}e^{\theta_{\epsilon,1}}, \cdots, a_{1N}e^{\theta_{\epsilon, N}}),\, \hat {\bf b}=(a_{21}^{*}e^{-\theta_{\epsilon,1}^{*}}, \cdots, a_{2N}^{*}e^{-\theta_{\epsilon,N}^{*}}),\, \Omega=(\Omega_{(lj)})_{N\times N},\, \Omega_{lj}=(a_{1l}^{*}a_{1j}e^{\theta_{\epsilon,l}^{*}+\theta_{\epsilon,j}}+a_{2l}^{*}a_{2j}e^{-\theta_{\epsilon,l}^{*}
-\theta_{\epsilon,j}})/(k_j-k_l^{*}),\,$
$\theta_{\epsilon,j}=-ik_j \Big[x-\Big(\sum_{\ell=1}^n\alpha_{2\ell+1}(-4k_j^2)^{\ell}|4k_j^2|^{\epsilon_{2\ell+1}}\Big)t\Big]$
with $a_{ij}$'s are constants, where $\{\{k_j,\, k_l,\, -k_l^*\}|s_{22}^{(\mathbb{R})}(k_j)=s_{22}^{(\mathbb{R})}(k_l)=0,\, k_j\in i\mathbb{R}^+,\, k_l\in\mathbb{C}^{+}\backslash i\mathbb{R}^{+}\}$ is the discrete spectrum set.

For $N=1$, $k_1=i\eta\, (\eta\in\mathbb{R}^+)$, and $a_{11},\, a_{21}\in\mathbb{R}\!\setminus\!\{0\}$, we have the fractional one-soliton solution of Eq.~(\ref{fmkdv}) as
\bee\label{u1}
q_{\rm fHOmKdV,1}(x,t)=2\eta\,{\rm sgn}(a_{11}a_{21})\,{\rm sech}\!\left[2\eta x-2\eta\left(\sum_{\ell=1}^n\alpha_{2\ell+1}(2\eta^2)^{2\ell+2\epsilon_{2\ell+1}}\right)t
+\ln\left|\frac{a_{11}}{a_{21}}\right|\right],
 \ene
which admits the multi L\'evy indexes ($\epsilon_\ell,\, \ell=1,2,...,n$) and displays the anomalous dispersion. The wave velocity is related to the multi L\'evy indexes, given by $v_w=\sum_{\ell=1}^n\alpha_{2\ell+1}(2\eta^2)^{2\ell+2\epsilon_{2\ell+1}}.$

\v {\it Conclusions and discussions.}---In conclusion, we have used a simple idea to present the new types of integrable multi-L\'evy-index and mixed fractional higher-order NLS, cmKdV, and mKdV hierarchies, and studied their explicit forms via the completeness of squared eigenfunctions, and anomalous dispersion relations via their linearizations. We also give the formulae for the fractional $N$-soliton solutions. The used new idea can also be extended to other new multi-index and mixed integrable fractional nonlinear soliton hierarchies. These obtained results will be useful to display the super-dispersion transports of nonlinear waves in multi-L\'evy-index or mixed fractional nonlinear media.

\v \noindent {\bf Acknowledgments}
 The work was supported by the National Natural Science Foundation of China (No. 11925108).


\vspace{-0.1in}


\begin{thebibliography}{10}
\setlength{\itemsep}{-0.35mm}
\makeatletter

\bibitem{Gardner1967}
C.~S. Gardner, J.~M. Greene, M.~D. Kruskal, and R.~M. Miura, Method for solving the
  {K}orteweg-de {V}ries equation, Phys. Rev. Lett. 19 (1967) 1095-1097.


\bibitem{ist2} M. J. Ablowitz and P. A. Clarkson, {\em Solitons, Nonlinear Evolution Equations and Inverse Scattering} ( Cambridge Univeristy Press, Cambridge, 1991).

\bibitem{ist1} Y. S. Kivshar and  G. P. Agrawal, {\em  Optical Solitons: from Fibers to Photonic Crystals} (Academic Press, New York, 2013).
    
\bibitem{rev1} Y. V. Kartashov, G. E. Astrakharchik, B. A. Malomed, and L. Torner, Frontiers in multidimensional self-trapping of nonlinear fields and matter, Nat. Rev. Phys. 1 (2019) 185-197.

\bibitem{rev2} D. Mihalache, Localized structures in optical and matter-wave media: A selection of recent studies, Rom. Rep. Phys. 73 (2021) 403.







\bibitem{fc-book2} K.S. Miller and B. Ross, {\it An Introduction to the Fractional Calculus and Fractional Differential Equations}
  (John Wiley \& Sons Inc., New York, 1993).

\bibitem{fc-book} S. Das, {\it Introduction to Fractional Calculus} (Springer, Berlin, 2011).

\bibitem{prl87} M. F. Shlesinger, B. J. West, and J. Klafter, L\'evy dynamics of enhanced diffusion: Application to turbulence, Phys. Rev. Lett. 58 (1987) 1100.

\bibitem{west97} B. J. West, P. Grigolini, R. Metzler, and T. F. Nonnenmacher, Fractional diffusion and L\'evy stable processes, Phys. Rev. E 55 (1997) 99-106.

\bibitem{pr20} R. Metzler and J. Klafter, The random walk's guide to anomalous diffusion: A fractional dynamics approach,
Phys. Rep. 339 (2000) 1.

\bibitem{fc-rev} N. Laskin, Fractional quantum mechanics, Phys Rev E  62 (2000) 3135-3145.


\bibitem{long15} S. Longhi, Fractional Schr\"odinger equation in optics, Opt. Lett. 40 (2015) 1117.

\bibitem{barkai2012} D. A. Kessler and E. Barkai, Theory of fractional L\'evy kinetics for cold atoms diffusing in optical lattices, Phys. Rev. Lett. 108 (2012) 230602.

\bibitem{fc-num} K. Diethelm, N. J. Ford, and A. D. Freed, A predictor-corrector approach for the numerical solution of fractional differential equations, Nonlinear Dyn. 29 (2002) 3-22.

\bibitem{fls00} N. Laskin, Fractional Schr\"odinger equation, Phys. Rev. E 66 (2002) 056108.

\bibitem{fc-book-num} C. Li and F. Zeng, {\it Numerical Methods for Fractional Calculus} (CRC Press, Boca Raton, 2015).





\bibitem{wangprl2020} W. Wang and E. Barkai, Fractional advection-diffusion-asymmetry equation, Phys. Rev. Lett. 125 (2020) 240606.

\bibitem{boris21} B. A. Malomed, Optical solitons and vortices in fractional media: A mini-review of recent results,
Photonics 8 (2021) 353.

\bibitem{Riesz} M. Riesz, L'int\'egrale de Riemann-Liouville et le probl\'eme de Cauchy, Acta Math. 81 (1949) 1–222.

\bibitem{li19} M. Cai and C.P. Li, On Riesz derivative, Fractional Calculus Appl. Anal. 22 (2019) 287–301.


\bibitem{ab-prl22}  M. J. Ablowitz,  J. B. Been, and L. D. Carr, Fractional integrable nonlinear soliton equations, Phys. Rev. Lett. 128 (2022) 184101.

\bibitem{ab22}  M. J. Ablowitz,  J. B. Been, and L. D. Carr,  Integrable fractional modified Korteweg-de Vries, sine-Gordon,
and sinh-Gordon equations, arXiv: 2203.13755 (2022).

\bibitem{yanfnls-2022}W. Weng, M. Zhang, and Z. Yan, Dynamics of fractional $N$-soliton solutions with anomalous dispersions of integrable fractional higher-order nonlinear Schr\"odinger equations, arXiv:2208.04493 (2022).

\bibitem{yanfmkdv-2022} M. Zhang, W. Weng, and Z. Yan, Interactions of fractional $N$-solitons with anomalous dispersions for the integrable combined fractional higher-order mKdV hierarchy, arXiv:2208.04497 (2022).


\bibitem{nls1} V. E. Zakharov and A. B. Shabat, Exact theory of two-dimensional self-focusing and one-dimensional self-modulation of waves in nonlinear media, Sov. Phys. JETP  34 (1972) 62-69 [Zh. Eksp. Teor. Fiz. 61 (1971) 118-134].

\bibitem{hirota} R. Hirota, Exact envelope-soliton solutions of a nonlinear wave equation, J. Math. Phys. 14 (1973) 805.

\bibitem{lpd} M. Lakshmanan, K. Porsezian, M. Daniel
Effect of discreteness on the continuum limit of the Heisenberg spin chain, Phys. Lett. A, 133  (1988) 483-488.

\bibitem{hmkdv1} M. Ito, An extension of nonlinear evolution equations of the KdV (mKdV) type to higher orders,
 J. Phys. Soc. Jpn. 49 (1980) 771.


\bibitem{akns} M. J. Ablowitz, D. J. Kaup, A. C. Newell, and H. Segur, The inverse scattering transform-Fourier analysis for
nonlinear problems, Stud. Appl. Math. 53 (1974) 249-315.

\bibitem{Zhou1989} X.~Zhou, Direct and inverse scattering transforms with arbitrary spectral singularities, Commun. Pure Appl. Math. 42 (1989) 895-938.


\bibitem{kaup76} D. J. Kaup, Closure of the squared Zakharov-Shabat eigenstates, J. Math. Anal. Appl. 54 (1976) 849-864.

\bibitem{yan2022} Z. Yan, Dynamics of multi-soliton solutions of integrable multi-index and mixed fractional nonlinear soliton hierarchies, preprint (2022).

\end{thebibliography}
\end{document}